\documentclass[twocolumn, longbib]{aastex701}
\usepackage{natbib}
\usepackage{xspace}
\usepackage[version=4]{mhchem}
\usepackage{subcaption} 
\usepackage{upgreek}
\usepackage{graphicx} 
\usepackage{fancyhdr}
\usepackage{amssymb}
\usepackage{xcolor}
\usepackage{orcidlink}
\usepackage{amsmath}
\usepackage{hyperref}
\hypersetup{
    colorlinks=true,
    linkcolor=blue,
    filecolor=magenta,      
    urlcolor=cyan
    }
\usepackage{booktabs}
\usepackage[acronym]{glossaries}
\glsdisablehyper
\makeglossaries
\newacronym{SETI}{SETI}{the Search for Extraterrestrial Intelligence}
\newacronym{ETI}{ETI}{extraterrestrial intelligence}
\newacronym{DSN}{DSN}{Deep Space Network}
\newacronym{GBT}{GBT}{Green Bank Telescope}
\newacronym{ATA}{ATA}{Allen Telescope Array}
\newacronym{ETZ}{ETZ}{Earth Transit Zone}

\newcommand{\fsec}{\ensuremath{.\kern-3.25pt\arcsec\kern-1pt}}
\newcommand{\fmin}{\ensuremath{.\kern-3pt\arcmin}}
\newcommand{\fdeg}{\ensuremath{.\kern-3.75pt\arcdeg\kern-1.25pt}}

\newcommand{\PSUAA}{Department of Astronomy and Astrophysics, The Pennsylvania State University, University Park, PA 16802, USA}
\newcommand{\PSUCEHW}{Center for Exoplanets and Habitable Worlds, The Pennsylvania State University, University Park, PA 16802, USA}
\newcommand{\PSETI}{Penn State Extraterrestrial Intelligence Center, The Pennsylvania State University, University Park, PA 16802, USA}

\begin{document}

\title{Detecting Extraterrestrial Civilizations That Employ an Earth-level Deep Space Network}

\correspondingauthor{Pinchen Fan} 
\email{pinchen@psu.edu}

\author[0000-0003-3988-9022]{Pinchen Fan}
\email{pinchen@psu.edu}
\affiliation{\PSUAA}
\affiliation{\PSETI}
\affiliation{\PSUCEHW}

\author[0000-0001-6160-5888]{Jason T.~Wright}
\email{astrowright@gmail.com}
\affiliation{\PSUAA}
\affiliation{\PSETI}
\affiliation{\PSUCEHW}

\author{T.~Joseph~W.~Lazio}
\email{Joseph.Lazio@jpl.caltech.edu}
\affiliation{Jet Propulsion Laboratory, California Institute of Technology, 4800 Oak Grove Dr, Pasadena, CA  91009}

\begin{abstract}
A major aspect of the search for extraterrestrial intelligence (SETI) involves searching for electromagnetic transmissions from extraterrestrial sources, often using our own transmissions as a guide. Previous studies have suggested that humanity's most consistently detectable technosignatures were transmissions from our deep-space networks and interplanetary radar. In this study, we analyze NASA Deep Space Network logs to explore what strategies for selecting SETI targets and scheduling observations would enhance the chances of detecting such networks. Analyzing Deep Space Network uplink transmission logs over the last 20 yr, we find that these emissions were predominantly directed along the ecliptic plane, towards or directly away from the Sun, and towards other planets. The average duty cycle within the Earth Transit Zone is 20 times higher than that across all ecliptic latitudes. In the case of Mars, we find a species that is able to observe the Solar System for radio emission during an Earth-Mars conjunction in the past 20 yr would have had a 77\% chance of observing during one of our transmissions, a $4\times10^5$-fold increase over intercepting our Deep Space Network transmission versus a random observer at a random time. These findings quantify how SETI searches might benefit from prioritizing edge-on exoplanet systems and aligning observation windows with exoplanetary conjunctions or planet-planet occultations because they significantly improve the likelihood of intercepting transmissions from any civilizations employing deep-space networks similar to our own.

\end{abstract}


\section{Introduction}
\label{sec:intro}

The search for extraterrestrial life is arguably one of the most captivating and profound topics in astronomy and astrophysics. The search for signs of past or extant life, regardless of an organism's complexity or intelligence, is conducted through biosignatures, such as \ce{O2}, \ce{O3}, \ce{NH3}, and \ce{CH4} \citep{biosig}, which are signs of biological activity. 

Meanwhile, \gls{SETI}  relies on technology as a proxy for intelligence and looks for signs of past or present technology (technosignatures) as evidence of current or historic intelligent life. Technosignatures also provide a potentially more unambiguous and detectable approach than biosignatures \citep{case_techno}. Furthermore, a successful \gls{SETI} search would not only offer the first proof of life outside Earth but also demonstrate the existence of a technological species detectable across interstellar distances. This would significantly deepen our understanding of the prevalence, complexity, and intelligence of life in the Universe, much more than what could be achieved by biosignature detection. 

Two of the nine axes of merit for technosignature searches \citep{axes} favor technosignatures that are already produced by humanity (the ``Extrapolation" axis) and are observable with current technology (the ``Detectability" axis). As such, previous studies have proposed various kinds of radio technosignatures emitted by humankind, assessed their detectability, and have often described the search for these technosignatures as searching for ``Earth-level civilizations" \citep{dsn_planet, BL_DSN, saide2023simulation}. More recently, \citet{ede} provided the most comprehensive calculations to date of the detectability of human-generated technosignatures in 2024. The study found that the two most detectable technosignatures emitted by humankind in 2024 were ``intermittent, celestially targeted radio transmissions" and ``persistent, celestially targeted radio transmissions" \citep{ede}. 

Intermittent, celestially-targeted radio transmissions include intentional transmissions such as the Arecibo message \citep{arecibomessage1974} or planetary radar. However, these transmissions are highly sporadic in timing and directionality. By contrast, persistent, celestially-targeted radio transmissions include much more frequent transmissions between ground stations and deep-space spacecraft, such as those from the NASA \gls{DSN}.

Previous research has proposed eavesdropping on interplanetary electromagnetic transmissions from edge-on exoplanet systems \citep{Siemion_2013}. There have been a few surveys that search for electromagnetic transmissions during exoplanet conjunctions \citep{traas2021breakthrough, transit_radio} and planet-planet conjunctions \citep{ppo}. Past research has also proposed searching for \gls{ETI}'s deep-space communication systems as an \gls{SETI} strategy \citep{dsn_planet}. Studies have cataloged stars located along the paths of five interplanetary probes, suggesting that these stars might have previously been illuminated by transmissions from our \gls{DSN} \citep{BL_DSN}. These searches reasonably assumed that such strategies could help boost the probability of a successful \gls{SETI} search. Here, we will quantify this intuition.

In this paper, we analyze archival \gls{DSN} transmission logs. Our goal is to characterize the timing and directionality of NASA’s \gls{DSN} broadcasts and identify any patterns in their detectability from outside the solar system, which could be shared by analogous networks used by an \gls{ETI}. By understanding the patterns in our own deep-space networks, we aim to develop effective strategies for identifying analogous extraterrestrial networks using our current technology.

In Sec.~\ref{sec:methods}, we describe how we collected \gls{DSN} transmission logs and correlated each transmission with the appropriate spacecraft ephemerides. In Sec.~\ref{sec:results}, we discuss how we generated detailed transmission maps and examine the patterns of the transmissions in terms of timing and directionality. In Sec.~\ref{sec:discussions}, we evaluate the implications of these findings for future \gls{SETI} searches.

\section{Analyzing Transmissions from the Deep Space Network}
\label{sec:methods}

\subsection{The Deep Space Network}
\label{sec:methods-dsn}

The \gls{DSN} consists of three sites, Goldstone in California, Madrid in Spain, and Canberra in Australia, each equipped with a 70-meter antenna and multiple (currently three to five) 34-meter antennas \citep{dsn_tech}. All antennas can transmit in the ``X~band" (approximately 7.190\,GHz)\footnote{This frequency is in the C band according to the IEEE definitions \citep{IEEE_radio}, but the \gls{DSN} refers to this as ``X-band".}, while many antennas can transmit in the S~band (approximately 2.110\,GHz), and three antennas can transmit in the Ka~band (approximately 34\,GHz) \citep{dsn_handbook}. Due to the limited number of antennas and low transmitting power ($<1$\,kW), Ka-band transmissions are excluded from our analysis. 

For the 70-meter antennas, the half-power beamwidths of S-band and X-band transmissions are about 0\fdeg128 and 0\fdeg038, respectively (\gls{DSN} Document Number 810-005, 101 in~\citet{dsn_handbook}). For the 34-meter antennas, the half-power beamwidths of S-band and X-band transmissions are about 0\fdeg258 and 0\fdeg0777, respectively \citep[\gls{DSN} Document Number 810-005, 103 in][]{dsn_handbook}. Typically, the antennas transmit up to \hbox{20\,kW}, with some antennas currently capable of transmitting up to \hbox{80\,kW} in X-band. 

The distance~$d$ at which a \gls{DSN} transmission can be detected by a distant observer is given by the radiometer equation \citep{BL2017}:
\begin{equation}
    \label{eq:rad_dist}
    d = \sqrt{\frac{G_{\mathrm{TX}} P_{\mathrm{TX}}}{4\pi (\mathrm{S/N}_\mathrm{min})(\mathrm{SEFD})} \sqrt{\frac{\tau_{\mbox{obs}}}{\Delta\nu}}},
\end{equation}
where $G_{\mathrm{TX}}$ is the transmitting antenna's gain, $P_{\mathrm{TX}}$ is the transmission power, $\mathrm{S/N}_\mathrm{min}$ is the signal-to-noise ratio threshold for detection, $\mathrm{SEFD}$ is the system equivalent flux density (SEFD) of the detector, $\tau_{\mbox{obs}}$ is the observation duration, and $\Delta\nu$ is the channel bandwidth of the observation, assumed wider than or equal to the transmission bandwidth. 

For a typical \gls{DSN} transmission, assume a transmission in X-band using a 70-meter antenna, with a bandwidth of 1 kHz and an input power of 20 kW. \gls{DSN} transmissions are typically scheduled for spacecraft communication sessions that can last an average of 3 hours. From the \gls{DSN} transmission logs (in Sec.~\ref{sec:schedule}), we have determined that only approximately 0.03\% of \gls{DSN} transmissions to deep-space spacecraft (as defined in Sec.~\ref{sec:methods_target}) move more than one beamwidth within a 30-minute interval. For the purpose of estimating detectability, we adopt a 30-minute integration time for a hypothetical observation. With an antenna gain of $10^{7.323}$ \citep[73.23~dBi, DSN Document Number 810-005, 101 in][]{dsn_handbook} and utilizing the \gls{GBT} as the detector with an SEFD of around 20\,Jy in X-band \citep{gbt_handbook}, such a transmission would be detectable with an $\mathrm{S/N}_\mathrm{min}$ of 5 from a distance of approximately 7\,pc, given a 30-minute observation.

\subsection{Target Selection}
\label{sec:methods_target}

The \gls{DSN} is used to communicate with spacecraft at a broad range of distances from Earth. In this study, we classify spacecraft into three categories: near-Earth, lunar, and deep-space spacecraft.

Near-Earth spacecraft are those operating within 0.95 of the Earth-Moon distance from Earth. These spacecraft include low-Earth orbit spacecraft, geostationary satellites, and similar missions, most of which do not employ the \gls{DSN}. Given that many countries deploy spacecraft to this region, our analysis of \gls{DSN} transmissions cannot realistically represent humanity’s overall transmissions to near-Earth spacecraft. Furthermore, since these spacecraft are relatively close, transmissions to them will generally be of much lower power than the \gls{DSN} typically employs, resulting in lower detectability. Accordingly, we exclude all \gls{DSN} transmissions to near-Earth spacecraft from our analysis, acknowledging that humanity's transmissions to these spacecraft are substantial, and a more thorough future analysis should include them. 

We define lunar spacecraft as those positioned between 0.95 and 1.05 times the Earth-Moon distance. Although the actual transmission power may not match that used for deep-space missions, there are numerous lunar missions, prompting us to include them in our analysis. Nevertheless, many other countries also operate lunar missions, so our analysis of \gls{DSN} transmissions will not fully capture humanity’s collective transmissions to lunar spacecraft.

We define deep-space spacecraft as those traveling beyond 1.05 times the Earth-Moon distance from Earth. These include spacecraft bound for Mars and other planets, as well as telescopes positioned around the Sun-Earth Lagrange points. Since most deep-space missions are NASA-led, we expect that the analysis of \gls{DSN} transmissions to these spacecraft should accurately reflect humanity's strongest deep-space transmissions.

\subsection{Deep Space Network Schedules}
\label{sec:schedule}

We retrieved the publicly available \gls{DSN} schedules from JPL's ``SPS REST Query Wrapper".\footnote{\url{https://spsweb.fltops.jpl.nasa.gov/rest/Wrapper/Schedule/activity.html}} The transmissions recorded within these schedules span from 00:00:00 UTC on 2005 January 1 to 00:00:00 UTC on 2025 January 1, covering 92.5 antenna-yr over this 20 yr interval. Approximately 76\% (70 antenna-yr) of all transmissions were in X-band, and 19\% (17 antenna-yr) involved X-band transmissions using 70-meter antennas. Only 3.3\% (3 antenna-yr) were directed to lunar spacecraft, indicating that the bulk of our dataset comprises transmissions to deep-space spacecraft.

The schedules record four times, a ``start of activity" (\texttt{starttime}), ''beginning of track'' (\texttt{BOT}), ``end of track'' (\texttt{EOT}), and ``end of activity" (\texttt{endtime}).  The ``activity'' represents the entire time required, including pointing the antenna to the spacecraft, configuring the antenna, performing any required calibrations, and the like. The interval \texttt{BOT} to~\texttt{EOT} indicates when the antenna is intended to be transmitting to or receiving telemetry from a spacecraft. Accordingly, we used \texttt{BOT} and \texttt{EOT} to represent start and end times of each transmission. The \texttt{facility} lists the antenna numbers, which we used to identify transmissions sent via 70-meter antennas, which have numbers 14, 43, and 63, and 34-meter antennas, with other numbers. 

At the time of writing, the numbers for the operational \gls{DSN} 34-meter antennas are: 24, 25, and 26 at Goldstone; 34, 35, and 36 at Canberra; and 53, 54, 55, 56, and 65 at Madrid. Additionally, antennas 14 and 27 at Goldstone and antenna 45 at Canberra were in service at various times during the 20 yr study period. Therefore, we include transmissions from the last three antennas as well, since they strengthen our analysis.

The \texttt{projuser} lists the code names of the targets, which we decode using NASA's ``Current Mission Set"\footnote{\url{https://deepspace.jpl.nasa.gov/about/commitments-office/current-mission-set/}}, revealing the actual target names. Whether a transmission occurred in S-band or X-band could be determined from \texttt{equipmentlist}, where ``STXL" denotes transmissions in S-band and ``XTXL" denotes transmissions in X-band. Given the transmission frequency and dish size, we found the half-power beamwidth of each transmission as discussed in Sec.~\ref{sec:intro}. The total transmission times to each target in each dish size–frequency configuration are shown in Appx. \ref{sec:appx-targets}.

These schedules do not include the actual transmitted power for each individual transmission. Therefore, we are unable to estimate the detectability for individual transmissions. However, we were able to obtain a summary of specific transmitted powers over a small period of time, through private communication. Across most antennas, approximately 50\%-60\% were sent at around 20 kW, with the rest under 20 kW. For transmission towards Mars, we have transmission power data for 1 antenna-yr of transmissions, accounting for only 3\% of total transmissions to Mars ($\sim$34 antenna-yr) over the last 20 yr. Of this small subset of transmissions to Mars, around 33\% were sent at around 20\,kW, with the rest below 20\,kW.\footnote{Our experience with the DSN Now website \url{https://eyes.nasa.gov/apps/dsn-now/dsn.html} (which does provide the transmission power) anecdotally shows that transmissions to deep-space spacecraft are typically around 20 kW.}

\subsection{Spacecraft ephemerides}
\label{sec:ephemerides}

We obtained the ephemerides of each spacecraft using the NASA JPL Horizons Web Application\footnote{\url{https://ssd.jpl.nasa.gov/horizons/}}. Because the distances between \gls{DSN} sites exceed 6000\,km \citep{dsn_tech}, if antennas at different \gls{DSN} sites were to point at the same target 0.95 Earth-Moon distance away, their pointing direction would differ by $\gtrsim$1\arcdeg, due to parallax, much larger than the smallest beamwidth of 0\fdeg038. In contrast, antennas at the same \gls{DSN} site are relatively close ($\sim$10\,km) to each other, so the differences in pointing directions would be no more than 0\fdeg0016 for antennas at the same site. Therefore, for each target, we downloaded separate ephemerides with ``Observer Location" in the Horizons system set to each of the three \gls{DSN} sites (DSS-14, DSS-43, DSS-63). Then, for transmissions originating from different antennas in the same \gls{DSN} site, e.g. DSS-14, we use the ephemerides with that site (DSS-14 in this example) as the origin.

The ephemerides for lunar spacecraft were sampled every 20 minutes, and, for deep-space spacecraft, a 60-minute sampling interval was used. Then, we use the \texttt{interp1d} function within \texttt{scipy} \citep{scipy} to linearly interpolate the ephemerides. We divided each transmission into segments so that the center of each successive segment’s pointing direction is just within half a beamwidth of that transmission. We then used the start time of every segment to obtain the interpolated ephemerides.

We also retrieved ephemerides with a 60-minute time step for the Sun and all planets other than Earth and followed the same steps for ephemeris interpolation for mapping the transmissions later in Sec.~\ref{sec:methods:mapping}.

\subsection{Light-travel Time Adjustment}
\label{sec:methods:mapping}

In the NASA JPL Horizons System, the ephemerides are provided ``compensated for down-leg light-time delay aberration," meaning that they show the apparent position of the target at a given time, which corresponds to \gls{DSN} antenna pointings when they are receiving transmissions. 

To calculate the actual directions of transmission and the apparent positions of the planets from the perspective of a distant observer, we must make some adjustments.

\begin{figure}[h]
    \centering
    \includegraphics[width=1\linewidth]{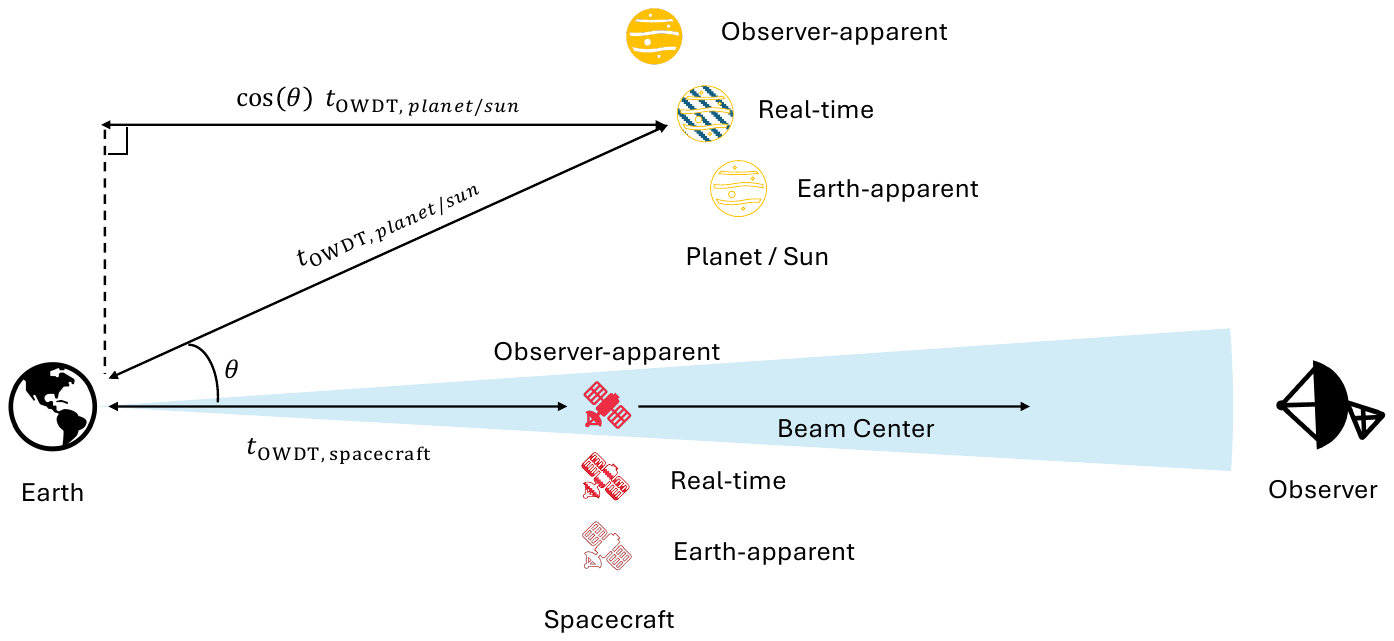}
    \caption{Illustration of the difference in apparent positions of the spacecraft and the planets/Sun, due to light-time delay, for Earth and for an extraterrestrial observer. Distances are expressed in terms of the one-way light-travel time (OWDT).}
    \label{fig:illustrate}
\end{figure}

The geometry of the problem is illustrated in Fig.~\ref{fig:illustrate}, which shows the positions of a spacecraft illuminated by the \gls{DSN} and a planet as they appear from Earth at the moment of transmission $t_{\rm transmit}$ (``Earth-apparent"), where they truly are at that moment (``real-time"), and at the positions they would appear to a distant observer upon receiving the \gls{DSN} signal (``observer-apparent").

To calculate these objects' real-time positions, we advanced the Earth-apparent positions by the interpolated one-way light time $t_\mathrm{OWDT}$, given by \texttt{1-one\_way\_DT} at $t_{\rm transmit}$. 

Since the \gls{DSN} antennas need to point at the future position of the spacecraft at the time the transmissions arrive at the spacecraft, we would need to find the ephemerides of the spacecraft by advancing the real-time ephemerides by $t_\mathrm{OWDT, spacecraft}$. Thus, we accomplish this by querying the Earth-apparent positions of the objects at a time $t_\mathrm{pointing}$
\begin{equation}
    t_\mathrm{pointing} = t_\mathrm{transmit}+2~t_\mathrm{OWDT, spacecraft}.
\end{equation}

This gives us the \textit{pointing directions} of every transmission. Then, we plotted the pointing directions in equatorial and ecliptic coordinates to reveal the pattern of past \gls{DSN} transmissions. This was particularly useful for visually identifying any regions of the sky with high transmission density.

When a distant observer detects a \gls{DSN} transmission, it would see Earth right behind the target spacecraft of the transmission. This Earth-spacecraft-observer line is simply defined by the pointing direction of that transmission. Then, to get the angle between the Sun or the planets and the pointing direction (which should be zero when we are transmitting to the Sun or the said planets), we need to find the positions of the Sun and the planets as they would appear to the observer. This requires an adjustment to the ``real-time" ephemerides by further compensating for the light-time delay from the Sun and the planets to the observer. This adjustment is $\cos{(\theta)}~ t_\mathrm{OWDT, sun/planet}$, where $\theta$ is the angle between the real-time position of the Sun or the planets and the pointing direction, and $t_\mathrm{OWDT, sun/planet}$ is the interpolated one-way light time from the Sun or the planets to Earth. Thus, to get the observer-apparent positions of the Sun and the planets, we query their ephemerides at the time~$t_\mathrm{obs-app}$, defined as:
\begin{equation}
 t_\mathrm{obs-app} = t_\mathrm{transmit} + \left(1+\cos{(\theta)}\right)~t_\mathrm{OWDT, Sun/planet}.
  \label{eqn:obs}
\end{equation}

\section{Results}
\label{sec:results}

From 00:00:00 UTC on 2005 January 1 to 00:00:00 UTC on 2025 January 1, \gls{DSN} transmissions have covered approximately 15\% of the celestial sphere. In this section, we generate detailed maps of the transmissions to investigate if there are any patterns in the transmissions.

We show a few selected figures in this paper. All figures produced in this section appear in the attached ``figure-set.zip" file. 

\subsection{Mapping DSN Transmission}
\label{sec:results-maps}

To show the general directional patterns in the \gls{DSN} transmissions, we map the pointing directions of the transmissions in both equatorial and true geocentric ecliptic coordinates. 

Then, in ecliptic coordinates, we also calculated the ``Sun angle ecliptic longitude" and ``Sun angle ecliptic latitude" as the difference in true geocentric ecliptic longitude or latitude between each pointing and the observer-apparent Sun, and we used similar terms for the planets. This shows whether past \gls{DSN} transmissions predominantly align with particular longitudinal or latitudinal angles with respect to the Sun or other planets. 

Instead of using the total time of transmissions, we plot the duty cycle of the \gls{DSN}: that is, the fraction of time during which there is at least one active transmission in a given direction. In calculating the duty cycle, we determined if at least one antenna was transmitting toward a pixel on the sky. Particularly for directions toward Mars, it is common, if not regular, for two or more antennas to be transmitting at the same time. Thus, we consider a pixel illuminated by two antennas to be the same as that pixel illuminated by only one antenna, so the maximum duty cycle for a given pixel is 100\%.

To generate these maps, we used \textsc{healpy} \citep{healpy} and \textsc{HEALPix} \citep{healpix} with $N_{\rm side}=4096$. The pixel size is around 0\fmin85, half the smallest \gls{DSN} beamwidth.  We then recorded the BOTs and EOTs for transmissions targeted at these pixels. We consider a pixel to be illuminated by a DSN transmission if any part of it falls within one-half half-power beamwidth of the beam's pointing direction. For each transmission, we used the \texttt{query\_disc} function in \textsc{healpy} (with \texttt{inclusive=True}) to identify the set of pixels illuminated by each DSN beam. Finally, we computed the duty cycle for each pixel using the compiled lists of BOTs and EOTs.

\begin{figure*}[ht!]
    \centering
    \includegraphics[width=1\linewidth]{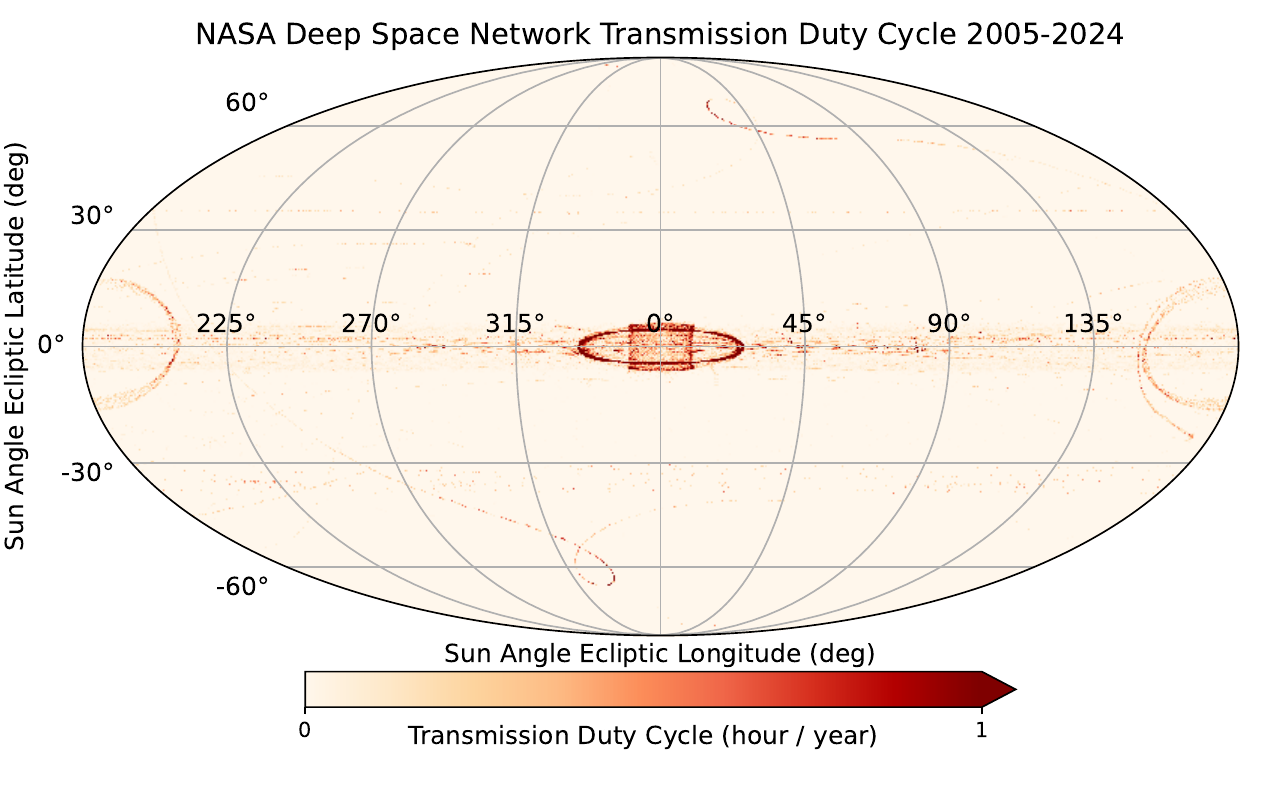}
    \caption{Map of DSN duty cycles using transmissions from 2005 to 2024. Around a Sun Angle Ecliptic Longitude of~0\arcdeg: the box-shaped area represents transmissions to the Advanced Composition Explorer (ACE) and the Deep Space Climate Observatory (DSCOVR); the edge of the oval-shape area represents transmissions to the Solar and Heliospheric Observatory (SOHO); the inside of that oval-shaped area represents transmissions to Wind. Around a Sun Angle Ecliptic Longitude of~180\arcdeg, the circle represents transmissions to the James Webb Space Telescope (JWST). The maps of DSN duty cycles across all coordinate systems (10 images) are available in the online journal.}
    \label{fig:sun_map}
\end{figure*}

Figure~\ref{fig:sun_map} shows the accumulated duty cycle of \gls{DSN} transmissions in terms of the angular distance between each transmitted beam and the Sun, expressed in the difference in their ecliptic coordinates. As viewed from a distant vantage point, the Sun-angle ecliptic longitude reflects the difference between the ecliptic longitudes of each transmitted beam and the Sun. The Sun-angle ecliptic latitude reflects the difference between the ecliptic latitudes of each transmitted beam and the Sun. In other words, for an outside observer, the Sun-angle ecliptic longitude reflects Earth’s orbital phase, while the Sun-angle ecliptic latitude is 90\arcdeg~minus Earth’s orbital inclination with respect to that observer's sky plane. 

Figure~\ref{fig:sun_map} shows dense transmissions around Sun-angle ecliptic longitudes of 0\arcdeg~and 180\arcdeg~(i.e., roughly in line with or opposite the Sun) while near the ecliptic plane. These regions correspond to transmissions toward spacecraft around the Sun-Earth Lagrange points L1 and L2. For example, the cluster around 180\arcdeg~in Sun-angle ecliptic longitude and 0\arcdeg~in Sun-angle ecliptic latitude includes transmissions to JWST, which librates about L2.

\subsection{Duty Cycle}
\label{sec:results-hist}

We plot the duty cycle against each coordinate discussed in Sec.~\ref{sec:results-maps} and the total angle between the pointing directions and the Sun or the planets. With this, we can determine if there is any special angle that the \gls{DSN} frequently transmits to. 

First, we set 1\arcmin-wide bins in each of the coordinates: longitude, latitude, and total angle, and calculate the average duty cycle across all pixels inside the 1\arcmin-wide bins.  Then, we plot the distribution of the average duty cycle. We also compare the duty cycle distribution of all available data (2005--2024) with a subset of more recent data (2022--2024), to inspect any changes in \gls{DSN} usage patterns.

\begin{figure}
    \centering
    \includegraphics[width=1\linewidth]{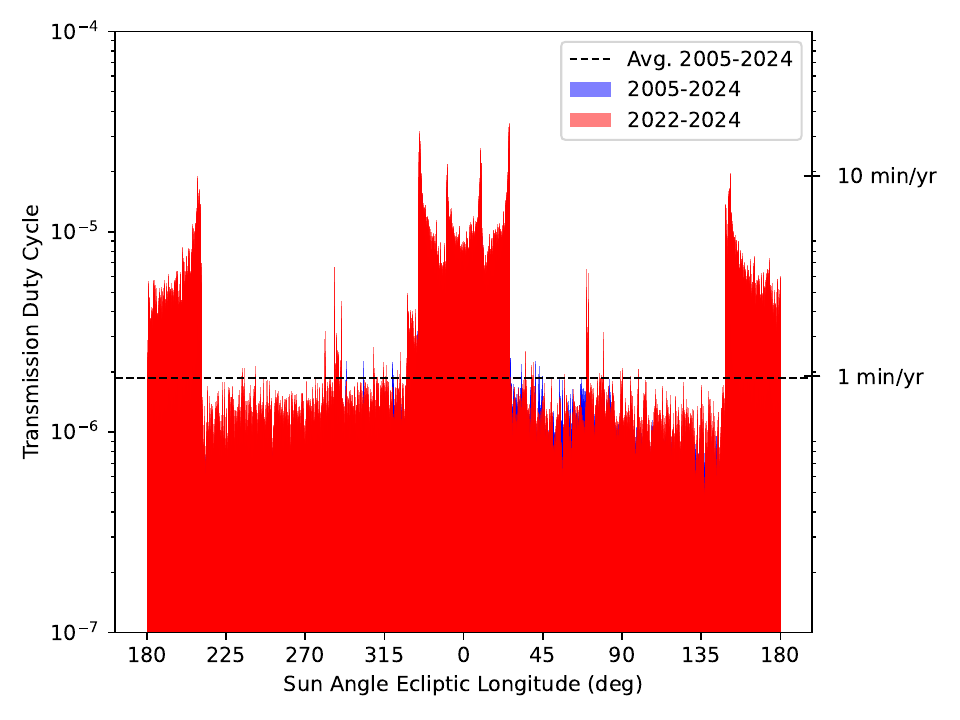}
    \caption{Distribution of DSN transmission duty cycle across Sun Angle Ecliptic Longitude. The average DSN transmission duty cycle among the pixels within each 1\arcmin-wide bin of Sun Angle Ecliptic Longitude is shown with 2005-2024 and 2022-2024 data. Transmissions to $\pm30\arcdeg$ of Sun angle ecliptic longitudes of 0\arcdeg~and 180\arcdeg~are about four times more frequent than transmissions to other Sun Angle Ecliptic Longitudes. The distributions of DSN duty cycles across all longitudes (10 images) are available in the online journal.}
    \label{fig:sun_lon}
\end{figure}

Figure~\ref{fig:sun_lon} offers a closer look at the duty cycle distribution in terms of Sun-angle ecliptic longitude. It reveals a broad peak within about $0\arcdeg\pm30\arcdeg$ in Sun-angle ecliptic longitude that has been persistent for the last 20\,yr, and a notable 4-fold increase around~180\arcdeg~in Sun-angle ecliptic longitude in the 2022--2024 data, likely due to transmissions to \textit{JWST}.

\begin{figure}
    \centering
    \includegraphics[width=1\linewidth]{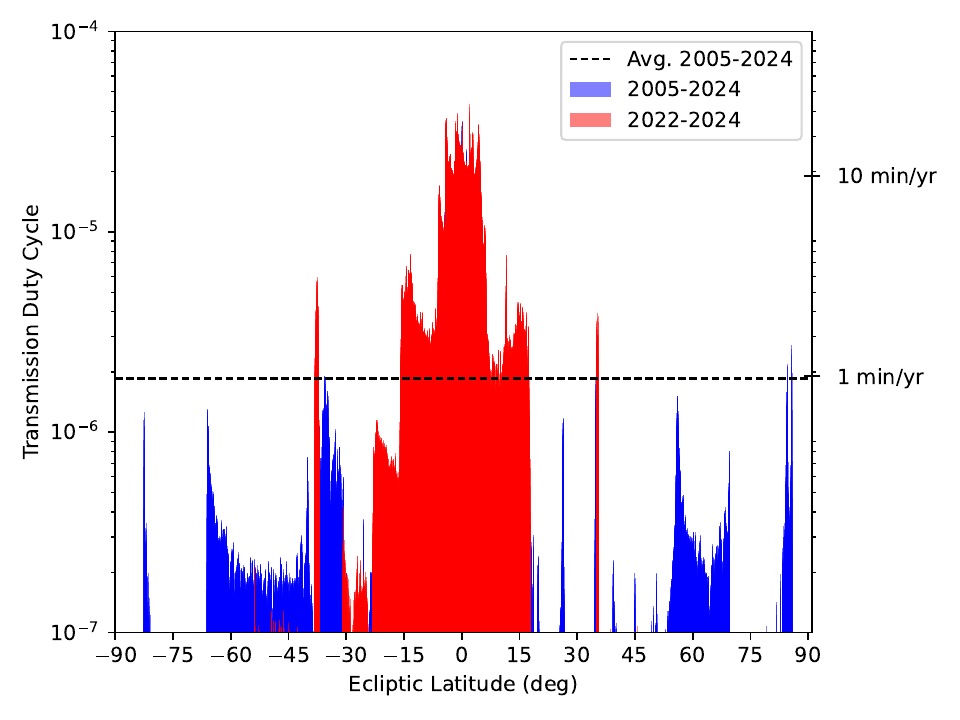}
    \caption{Distribution of DSN transmission duty cycle across Ecliptic Latitude. The average DSN transmission duty cycle among the pixels within each 1\arcmin-wide bin of Ecliptic Latitude is shown with 2005-2024 and 2022-2024 data. Around 79\% of all transmissions, in terms of total transmission duration, were sent to within 5\arcdeg~of an ecliptic latitude of 0\arcdeg. The average duty cycle within $\sim$0\fdeg27 of the ecliptic plane (the size of the \gls{ETZ}) is 20 times the average duty cycle across all ecliptic latitudes. The distributions of DSN duty cycles across all latitudes (10 images) are available in the online journal.}
    \label{fig:ecl_lat}
\end{figure}

Figure \ref{fig:ecl_lat} shows the duty cycle distribution across ecliptic latitude. It exhibits a pronounced peak in the DSN transmission duty cycle near the ecliptic plane. In ecliptic coordinates, 84\% of the pixels within $\pm5\arcdeg$ of the ecliptic plane were 
illuminated at least once by a DSN transmitter in the 20 yr between 2005 and 2024, compared to less than 3\% of pixels outside that region. The total transmission duration for all pixels within $\pm5\arcdeg$ of the ecliptic plane ($\sim$73 antenna-yr) accounts for 79\% of the cumulative DSN transmission time over those 20 yr (92.5 antenna-yr). This concentration indicates that most DSN transmissions were sent within a few degrees of the ecliptic plane, consistent with the fact that most interplanetary spacecraft operate near the ecliptic. The average duty cycle for pixels within $\sim$0\fdeg27 of the ecliptic plane, the width of the \gls{ETZ}, is about 13 minutes of transmission per year, roughly 20 times higher than the average duty cycle of 0.65 minutes per year across all 1\arcmin-wide ecliptic latitudes bins.

\begin{figure}
    \centering
    \includegraphics[width=1\linewidth]{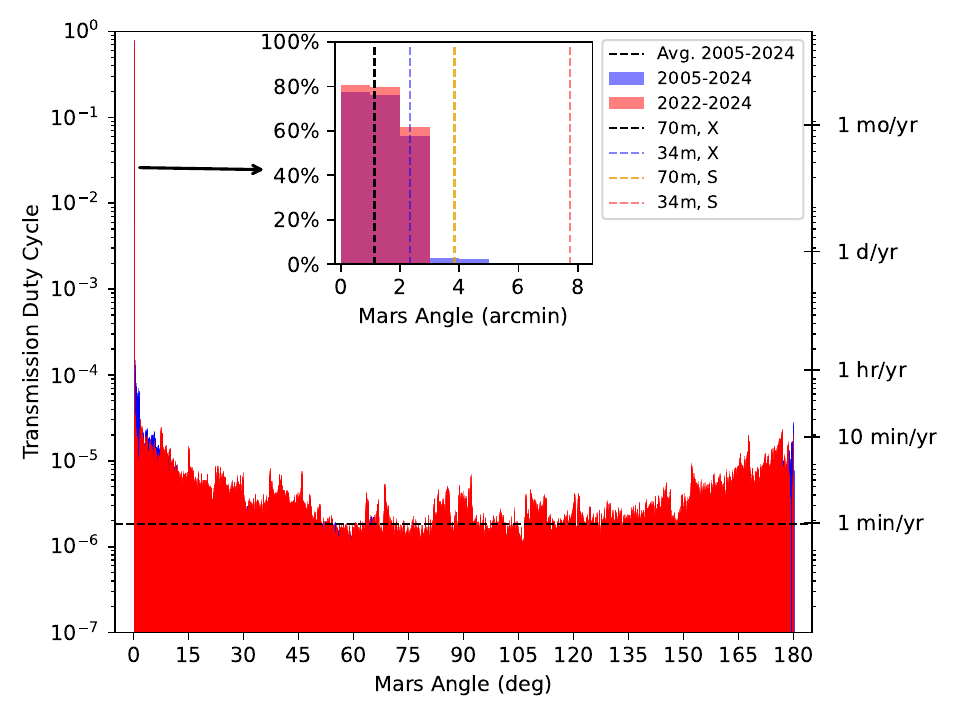}
    \caption{Distribution of DSN transmission duty cycle as a function of angular separation from Mars. The average DSN transmission duty cycle among the pixels within each 1\arcmin-wide bin of total Mars Angle is shown with 2005-2024 and 2022-2024 data. The inset histogram emphasizes the distribution within 8\arcmin~of Mars; dashed vertical lines mark the \gls{DSN} beam radii, with dish diameter and transmission frequency given in the legend. Over the past 20 yr, \gls{DSN} has transmitted to within 2\arcmin~of Mars 77\% of the time, and to between 2\arcmin~and 3\arcmin~of Mars 58\% of the time. Within 2\arcmin~of Mars, the duty cycle is $4\times10^5$ times the all-sky average. The distributions of DSN duty cycles across total angles from the Sun and the other planets (8 images) are available in the online journal.}
    \label{fig:mars_lon}
\end{figure}

NASA and other international space agencies have sent many spacecraft to Mars in the past 20 yr, and Fig.~\ref{fig:mars_lon} shows the duty cycle distribution in terms of Mars angle. The duty cycle peaks at about 77\% (9.4 months per year) within 2\arcmin~and slightly drops to around 58\% (7 months per year) for the region between 2\arcmin~and 3\arcmin~from Mars. The duty cycle quickly drops outside of 3\arcmin~from Mars. Within 2\arcmin~of Mars, the duty cycle is $4\times10^5$ times higher than the average duty cycle across the celestial sphere. This peak is prominent in both 2005--2024 and 2022--2024 data, reflecting humanity's long-standing focus on Mars exploration. Weaker but still visible peaks arise for Mercury, Jupiter, and Saturn (see the figure set available in the online journal), though they vary more over time. The average duty cycle within 2\arcmin~ of the line connecting Earth and a planet is $7\times10^4$ times higher than the average duty cycle across the celestial sphere. 

\section{Impact on Future SETI Surveys}
\label{sec:discussions}

Our findings suggest several strategies for enhancing the effectiveness of future \gls{SETI} searches.

First, our study helps prioritize the target list. We show that humanity's most powerful transmissions center heavily around a few degrees of the ecliptic plane. Most notably, in terms of total transmission duration, around 79\% of past \gls{DSN} transmissions were directed within 5\arcdeg~of the ecliptic plane. Although the duty cycle within the \gls{ETZ} is still only 13 minutes per year, it is 20 times higher than the average duty cycle across all ecliptic latitudes. To search for a similar deep-space network from an exoplanet system, it would be advantageous to choose systems hosting high-inclination exoplanets, such as transiting systems. Since, for most planetary systems, planets' orbits are close to co-planar, the presence of one or more transiting exoplanet(s) implies (but does not guarantee) a similarly high orbital inclination for other potential planets in the system. This increases the odds that a \gls{DSN}-like transmission beam would intersect Earth, by 20 times if we assume a \gls{DSN} analog. This strongly validates the strategies of \citep{traas2021breakthrough, franz2022breakthrough, transit_radio} to observe transiting planetary systems.

Second, our results show there may be preferred epochs for detecting \gls{DSN}-like transmissions. Our data show that persistent transmissions to Mars lead to a 77\% duty cycle within 2$^\prime$~of Mars, more than $4\times10^5$ times the average duty cycle across the celestial sphere. Searching during planet-planet occultations, as demonstrated by \citet{ppo}, could therefore amplify detection prospects significantly, in the Earth-Mars case, by a factor of potentially as much as $4\times10^5$. Of course, this approach relies on having knowledge of multiple exoplanet orbits in a system. In cases where there is only one known exoplanet in a system, we would need to rely on its orbital phase (which, in our analogy with \gls{DSN} transmissions, corresponds to the ecliptic longitude Sun angle of our transmissions). In recent years, \gls{DSN} transmissions have peaked around Sun angle ecliptic longitudes of 0\arcdeg~and 180\arcdeg, driven by telescope missions at Sun-Earth L1 and L2 (e.g., \textit{JWST}). If an \gls{ETI} similarly placed telescopes at their planet's own Lagrange points, transmissions could peak during its planet-star conjunctions. This might roughly double the chance of detection under an Earth-analog \gls{DSN} model, such as the search done in \citet{ETZ2}.

Third, the region within about 5\arcdeg~of our ecliptic plane received 79\% of all \gls{DSN} transmissions in terms of total transmission duration. As human space exploration expands, especially if long-term projects like human settlements on Mars \citep{beech2021terraforming} become a reality, communications may remain concentrated around the ecliptic plane. Over time, star systems located along this band may be most likely to receive and recognize our signals, potentially prompting a reply. Hence, targeting nearby stars within 5\arcdeg~of Earth’s ecliptic plane might be advantageous for future \gls{SETI} observations. This region includes the \gls{ETZ} \citep{ETZ1}. In Sec.~\ref{sec:methods-dsn}, we show that our technology can detect \gls{DSN} transmissions from 7\,pc away. There are 128 known star and/or brown dwarf systems located within 7\,pc of the Sun \citep[with parallaxes exceeding 142.86\,mas,][]{gaia}. If there were a civilization with technology at or above our level around any of those 128 stars, they may be able to detect us. Likewise, we could similarly detect \gls{DSN}-like transmissions originating from an \gls{ETI} located within these 128 systems. There have been searches for radio technosignatures from stars within the \gls{ETZ} \citep{ETZ2}. Future \gls{SETI} searches can expand to nearby stars within 5\arcdeg~of Earth's ecliptic plane to search for potential reply messages since these stars would likely be along the paths of past \gls{DSN} transmissions. \citet{meti} determined the earliest possible reply time for intentional transmissions to extraterrestrial intelligence. Future work could identify nearby stars targeted by past \gls{DSN} transmissions and estimate the earliest possible dates for replies.

Although our analysis focuses on \gls{DSN} radio transmissions, a similar pattern would apply to other communication methods, such as optical lasers. As NASA and other agencies develop prototype optical deep-space communication networks \citep{dsn_optical_1, dsn_optical_2, dsn_optical_3}, the pointing geometry is likely to remain the same. We note that laser beams could be extremely tight, for example, a 0\fsec2 beamwidth for a 1-micron laser emission produced by a diffraction-limited 1-meter transmitter. Since this is $\sim$$4\times10^4$ times tighter than the smallest \gls{DSN} beam, less spillover would be expected in optical communications. This means that fewer stars will be within a beam, and the beam might be blocked by a planet in the case of spacecraft orbiting or on the surface of that planet, and so not reach interstellar space. Nevertheless, if some spillover occurred from an optical deep-space network, the strategies discussed above would likely still be optimal search candidates to look for optical signals.

Our results show when an \gls{ETI} would be \textit{illuminated} by a \gls{DSN} beam, rather than if they would successfully \textit{detect} it with their instrument. As discussed in Sec.~\ref{sec:schedule}, the publicly available \gls{DSN} transmission logs do not include the actual transmission power. Therefore, our analysis focuses primarily on the directional patterns of these transmissions. Equation \ref{eq:rad_dist} shows that a typical (i.e. not the most powerful) \gls{DSN} transmission could be detected up to 7\,pc away. Previous studies also suggest that \gls{DSN} transmissions are detectable with current Earth-based technology at distances of up to several tens of parsecs under various detector configurations \citep{dsn_planet, BL_DSN, ede}. An \gls{ETI} might possess a more powerful deep-space communication system, making their signals more detectable by us. Additionally, an \gls{ETI} with more advanced detection capabilities could potentially detect our \gls{DSN} transmissions from even greater distances.

Furthermore, we do not include lower-power transmissions, such as non-\gls{DSN} ground stations. However, such transmissions are typically less detectable over interstellar distances. Since it would be extremely difficult to compile all transmissions to low-Earth-orbit satellites, one may need to simulate transmissions to a cloud of low-Earth-orbit spacecraft as ``radio clamor" and estimate its detectability, similar to the simulations on radio leakage from mobile towers \citep{saide2023simulation}. We do not expect any significant preference in the timing or directionality of these transmissions, given the wide range of orbits available for low-Earth-orbit spacecraft. 

Transmissions to satellites in or near geosynchronous orbit may show greater directionality and employ higher transmission power compared to low-Earth-orbit radio clamor, due to their greater distance and more restricted declination. Nevertheless, these transmissions likely have much lower power than \gls{DSN} transmissions. A detailed analysis of uplink transmissions targeting geosynchronous satellites can be explored in future studies.

Furthermore, some transmissions to lunar probes may be intercepted by the Moon completely, but only 3\% of all transmissions are to lunar spacecraft, so this should not affect our analysis significantly.

Future research may also try to retrieve transmission logs from deep-space networks built by other space agencies to improve our characterization of the pattern of all of humanity's deep-space transmissions. For example, China's Deep Space TT\&C (Tracking, Telemetry, and Command) system \citep{china_dsn} is able to transmit at 18 kW in S-band and 15 kW in X-band. Even though we only analyzed \gls{DSN} transmission logs, other countries' space programs may share similar goals, such as exploring planets in the solar system \citep{china_space_program, china_space_program_2}. Thus, the pattern of their transmissions may be similar to that of \gls{DSN}.

\section{Conclusion}
\label{sec:conclude}

\citet{cocconi1959searching} catalyzed modern \gls{SETI} with the insight, shared by Frank Drake \citep{drake1,drake2} and others, that humanity was capable of detecting its own radio transmissions at interstellar distances if those transmissions were sufficiently powerful and targeted. This has justified radio \gls{SETI} for decades because ``[radio SETI] assumes nothing about other civilizations that has not transpired in ours" \citep{sagan}. Here, we have shown that humanity has already reached that threshold, particularly for stars near the ecliptic plane (thus including the \gls{ETZ}), especially for those that witness close Earth-Mars conjunctions.

Novel to this work, we have analyzed actual transmission logs from the \gls{DSN} to quantify the directional patterns of its uplink signals. We accomplished this by mapping the duration of transmissions in multiple coordinate systems: equatorial, ecliptic, and angles relative to the Sun and other planets. We also plotted transmission duty cycles to quantify how frequently each region received \gls{DSN} signals.

Our analysis reveals a significant concentration of \gls{DSN} transmissions around Sun-Earth L1 and L2, as well as near planetary targets, with 79\% of transmissions confined within 5\arcdeg~of the ecliptic plane. Within the \gls{ETZ}, the duty cycle is 20 times higher than the average duty cycle across all ecliptic latitudes. This pattern suggests that exoplanet systems with nearly edge-on orbits, especially those hosting transiting planets, could be prime targets for searches aiming to detect \gls{DSN}-like technosignatures. Furthermore, focusing observations near exoplanet conjunctions or planet-planet occultations might increase the likelihood of intercepting transmissions from any similarly oriented deep-space network. In the case of Mars, persistent \gls{DSN} transmissions to Mars lead to over 77\% duty cycle within 2\arcmin~of Mars, $4\times10^5$ times the average duty cycle across the celestial sphere. 

When an external observer sees the Earth behind another planet, their chance of detecting a \gls{DSN} transmission rises to about 12\%, over $6\times10^4$ times higher than if they were observing at a random time.

As humankind broadens its presence in the Solar System, sustained transmissions to other planets would further intensify the concentration of strong transmissions along the ecliptic plane. Our analysis underscores the value of using humanity’s own \gls{DSN} as a model for how an early-phase spacefaring civilization might transmit, and how we might detect it.

\begin{acknowledgments}

A part of this work was funded by the NASA Exoplanets Research Program grant 80NSSC25K7145. The Penn State Extraterrestrial Intelligence Center and the Center for Exoplanets and Habitable Worlds are supported by the Pennsylvania State University and the Eberly College of Science. Computations for this research were performed on Penn State’s Institute for Computational and Data Sciences’ Roar supercomputer. A part of this research was carried out at the Jet Propulsion Laboratory, California Institute of Technology, under a contract with the National Aeronautics and Space Administration. Some of the results in this paper have been derived using the \texttt{healpy} and \texttt{HEALPix} package. This research has made use of the Astrophysics Data System, funded by NASA under Cooperative Agreement 80NSSC21M00561. We thank M.~Johnston and B.~Arroyo for explaining the intricacies of the \gls{DSN} schedule. We thank F.~Huang and J.~Liao for assistance with obtaining the \gls{DSN} transmitted power levels. We also thank the anonymous reviewer for their constructive comments. 

\end{acknowledgments}

\appendix

\section{Transmission Targets}
\label{sec:appx-targets}

Tables \ref{tab:projuser-1} and \ref{tab:projuser-2} show the total transmission duration from 00:00:00 UTC on 2005 January 1 to 00:00:00 UTC on 2025 January 1 for every \gls{DSN} target (\texttt{projuser}) included in our dataset, and the associated mission name for each \texttt{projuser} code.

\begin{table}[h!]
\centering
\begin{tabular}{|c|c|c|c|c|c|}
\hline
\texttt{projuser} & Mission Name & \multicolumn{4}{c|}{Transmission Duration (antenna-hour)} \\
\hline
 &  & 34, S & 34, X & 70, S & 70, X \\
\hline
ACE & Advanced Composition Explorer & 22345 & 0 & 1 & 0 \\
\hline
APM1 & Peregrine & 0 & 74 & 0 & 0 \\
\hline
BEPI & BepiColombo & 0 & 209 & 0 & 52 \\
\hline
BIOS & BioSentinel & 0 & 412 & 0 & 62 \\
\hline
CAPS & CAPSTONE  & 0 & 2169 & 0 & 0 \\
\hline
CAS & Cassini  & 4 & 20030 & 0 & 13686 \\
\hline
CH2L & Chandrayaan-2 (Lander) & 107 & 0 & 0 & 0 \\
\hline
CH2O & Chandrayaan-2 (Orbiter) & 1343 & 0 & 0 & 0 \\
\hline
CH3 & Chandrayaan-3 (Lander spacecraft) & 220 & 0 & 0 & 0 \\
\hline
CH3P & Chandrayaan-3P (Orbiter) & 9 & 0 & 0 & 0 \\
\hline
DART & Double Asteroid Redirection Test & 0 & 1784 & 0 & 256 \\
\hline
DAWN & Dawn & 0 & 24028 & 0 & 5318 \\
\hline
DIF & Deep Impact Flyby & 0 & 7549 & 0 & 879 \\
\hline
DSCO & Deep Space Climate Observatory & 1141 & 0 & 0 & 0 \\
\hline
EM1 & Artemis I  & 391 & 0 & 0 & 0 \\
\hline
EMM & Emirates Mars Mission & 0 & 4015 & 0 & 458 \\
\hline
EQUL & EQUULEUS  & 0 & 6 & 0 & 0 \\
\hline
EURC & Europa Clipper & 0 & 1256 & 0 & 332 \\
\hline
GAIA & Gaia & 0 & 933 & 0 & 0 \\
\hline
GRLA & GRAIL-A & 5779 & 0 & 2 & 0 \\
\hline
GRLB & GRAIL-B & 5799 & 0 & 0 & 0 \\
\hline
HYB2 & Hayabusa 2 & 0 & 3195 & 0 & 271 \\
\hline
ICE & International Cometary Explorer & 5 & 0 & 0 & 0 \\
\hline
JNO & Juno  & 0 & 24720 & 0 & 14544 \\
\hline
JWST & James Webb Space Telescope & 14643 & 0 & 15 & 0 \\
\hline
KEPL & Kepler & 0 & 10677 & 0 & 1629 \\
\hline
KPLO & Korea Pathfinder Lunar Orbiter & 8066 & 0 & 31 & 3 \\
\hline
LADE & LADEE & 2346 & 0 & 0 & 0 \\
\hline
LCRO & LCROSS Centaur Impactor & 7 & 0 & 0 & 0 \\
\hline
LFL & Lunar Flashlight & 0 & 981 & 0 & 0 \\
\hline
LICI & LICIACube & 0 & 240 & 0 & 48 \\
\hline
LRO & Lunar Reconnaissance Orbiter & 4084 & 0 & 0 & 0 \\
\hline
LUCY & Lucy  & 0 & 5039 & 0 & 759 \\
\hline
M01O & Mars Odyssey & 0 & 70956 & 0 & 15498 \\
\hline
M20 & Mars2020  & 0 & 8509 & 0 & 2562 \\
\hline
\end{tabular}
\caption{Total transmission duration towards included \gls{DSN} targets. \texttt{projuser} is the target code in the \gls{DSN} transmission logs. ``Mission Name" is the name of the mission associated with each \texttt{projuser} code. The transmission durations are expressed for each pair of dish diameter and transmission frequency.}
\label{tab:projuser-1}
\end{table}
\begin{table}[h!]
\centering
\begin{tabular}{|c|c|c|c|c|c|}
\hline
\texttt{projuser} & Mission Name & \multicolumn{4}{c|}{Transmission Duration (antenna-hour)} \\
\hline
 &  & 34, S & 34, X & 70, S & 70, X \\
\hline
MCOA & MarCO-A & 0 & 472 & 0 & 71 \\
\hline
MCOB & MarCO-B & 0 & 312 & 0 & 54 \\
\hline
MER1 & Opportunity  & 0 & 4054 & 0 & 700 \\
\hline
MER2 & Spirit  & 0 & 1753 & 0 & 283 \\
\hline
MEX & Mars Express  & 14 & 8830 & 14 & 1833 \\
\hline
MOM & Mars Orbiter Mission & 5460 & 0 & 1248 & 0 \\
\hline
MRO & Mars Reconnaissance Orbiter & 0 & 90047 & 0 & 28644 \\
\hline
MSGR & MESSENGER & 0 & 20291 & 0 & 3254 \\
\hline
MSL & Curiosity & 0 & 13658 & 0 & 2816 \\
\hline
MUSC & Muses-C & 0 & 962 & 0 & 30 \\
\hline
MVN & MAVEN & 0 & 34294 & 0 & 2984 \\
\hline
NEAS & NEA Scout & 0 & 85 & 0 & 31 \\
\hline
NHPC & New Horizons & 0 & 5904 & 0 & 27917 \\
\hline
NSYT & InSight & 0 & 2903 & 0 & 533 \\
\hline
OMOT & OMOTENASHI  & 0 & 6 & 0 & 0 \\
\hline
ORX & OSIRIS-Rex & 0 & 21164 & 0 & 2008 \\
\hline
PLC & Akatsuki & 0 & 4246 & 0 & 713 \\
\hline
PSYC & Psyche  & 0 & 2137 & 0 & 203 \\
\hline
ROSE & Rosetta & 55 & 4824 & 111 & 2806 \\
\hline
SLIM & SLIM & 39 & 0 & 0 & 0 \\
\hline
SOHO & SOHO & 61470 & 0 & 1 & 0 \\
\hline
SPP & Parker Solar Probe & 0 & 14563 & 0 & 990 \\
\hline
STA & STEREO-A  & 0 & 24764 & 0 & 11939 \\
\hline
STB & STEREO-B & 0 & 15964 & 0 & 4081 \\
\hline
TESS & TESS & 252 & 0 & 0 & 0 \\
\hline
TGO & ExoMars Trace Gas Orbiter & 0 & 6000 & 0 & 748 \\
\hline
THB & THEMIS-B & 3762 & 0 & 0 & 0 \\
\hline
THC & THEMIS-C & 3483 & 0 & 0 & 0 \\
\hline
ULYS & Ulysses  & 7165 & 0 & 3332 & 0 \\
\hline
VEX & Venus Express  & 0 & 117 & 5 & 260 \\
\hline
VGR1 & Voyager 1 & 12 & 0 & 5074 & 4 \\
\hline
VGR2 & Voyager 2 & 2462 & 8 & 6119 & 4 \\
\hline
WIND & Wind & 17741 & 0 & 0 & 0 \\
\hline
\end{tabular}
\caption{Total transmission duration towards included DSN targets. \texttt{projuser} is the target code in the DSN transmission logs. ``Mission Name" is the name of the mission associated with each \texttt{projuser} code. The transmission durations are expressed for each pair of dish diameter and transmission frequency.}
\label{tab:projuser-2}
\end{table}

\bibliography{main}

\end{document}